\documentclass[10pt]{article}
\begin{document}

\title{Gravitational waveforms for finite mass binaries}
\author{M\'aty\'as Vas\'uth and J\'anos Maj\'ar
\footnote{Electronic addresses: {\tt vasuth@rmki.kfki.hu, majar@rmki.kfki.hu}} \\ 
{\em\small KFKI Research Institute for Particle and Nuclear Physics,}\\
{\em\small Budapest 114, P.O.Box 49, H-1525 Hungary}}

\maketitle

\begin{abstract}
\noindent 
One of the promising sources of gravitational radiation is a binary system composed of compact stars. It is an 
important question how the rotation of the bodies and the eccentricity of the orbit affect the detectable signal. 
Here we present a method to evaluate the gravitational wave polarization states for inspiralling compact binaries 
with comparable mass. We consider eccentric orbits and the spin-orbit contribution in the case of one spinning object 
up to 1.5 post-Newtonian order. For circular orbits our results are in agreement with existing calculations.
\end{abstract}

\maketitle

\section{Introduction}

During the inspiral and merger of compact stars intense gravitational radiation is produced, whose detection is 
expected by the current generation of gravitational wave observatories and more sensitive instruments, such as 
advanced LIGO\cite{advLIGO} and the LISA mission \cite{LISA}. The different characteristics of the binary, {\it e.g.} 
rotation of the bodies and eccentricity of the orbit contribute to the emitted radiation. Having accurate information 
about the waveforms will significantly increase the possibility of detection and the precision with which the source 
parameters can be identified.

The major approximation tool for the description of the dynamics and the generated waveforms of a binary system is 
the post-Newtonian (PN) expansion \cite{BDI}. Neglecting ultrarelativistic and extreme gravitational effects in this 
weak field approximation the velocities and the gravitational potential involved are small and there is no 
restriction on the mass ratio of the components. This approach precisely describes the motion of an inspiralling 
binary up to the last stable circular orbit \cite{Blanchet,BCV2}. To explore the evolution of the binary in the 
merger or ringdown phases one has to use different methods \cite{EOB} or numerical simulations to solve the Einstein 
equations \cite{Num}.

The polarization states of the emitted gravitational waves was computed for quasi-circular \cite{LW,BIWW,BDE} and 
elliptic orbits \cite{BS,GI,TG}. For spinning binaries the evaluation of the wave pattern has been done by several 
authors \cite{WW,Kidder,ACST}. In most cases the waveform is expressed formally in terms of the dynamical quantities 
of the motion. Here we give the expressions of the wave polarization states $h_{+}$ and $h_{\times}$ up to 1.5 
relative PN order for eccentric orbits. Extending our former results \cite{LT1} we discuss the effects of rotation 
for a binary system in which one of the bodies is spinning with spin vector ${\bf S}$ and the components have 
comparable masses $m_1$ and $m_2$.

In Ref.~\cite{GPV2} the equations of motion for a comparable mass binary system are derived by the use of the 
Lagrangian formalism and an appropriate radial parameterization of the orbit was given \cite{Param}. We use these 
results to determine the features of the motion up to 1.5\,PN order. In order to avoid long expressions we introduce 
the invariant and comoving coordinate systems which are fixed to the total and Newtonian angular momentum vectors, 
respectively. In Sec. 2 we evaluate the quantities which are necessary to obtain the general formulae of the 
detectable gravitational wave signals. In the first subsection we introduce the invariant and the comoving coordinate 
systems and give the formal expressions of the $({\bf N}$, ${\bf p}$, ${\bf q})$ triad which determines the relative 
orientation of the source and the observer. In the second subsection we describe the precession of the spin. In the 
third one, with the use of the constants of the motion, we give the equations for the radial and angular variables of 
the orbit linearly in spin. In Sec. 3 we recall that the detectable signal separates into $h_+$ and $h_{\times}$ and 
calculate the formal expressions of the polarization states from the transverse-traceless tensor $h^{ij}_{TT}$ which 
represents metric perturbations. Having a great importance, in Sec. 4 and 5 we investigate the circular orbit case 
and the extreme mass ratio limit. Our results for circular orbits are in agreement with those of Kidder 
\cite{Kidder}. Sec. 6 and the Appendix contain our conclusions and the general expressions for the 
transverse-traceless tensor in our case.

We use units in which $c=G=1$.

\section{Description of the motion}

\subsection{Basic vectors in the invariant and comoving coordinate systems}

To describe the time dependence of the orbital elements we introduce a coordinate system which does not change in 
time. The $z$ axis of this invariant system is fixed to the direction of the total angular momentum vector ${\bf J}$. 
This vector is constant up to 2\,PN order \cite{Kidder}. We choose the $x$ and $y$ axes in a way that the vector 
representing the direction of the line of sight has the form ${\bf N}=\left(\sin{\gamma},0,\cos{\gamma}\right)$ in 
this system, where $\gamma$ is the constant angle between ${\bf J}$ and ${\bf N}$.

As in the case of a Lense-Thirring system \cite{LT1} we introduce the comoving coordinate system, in which the $x$ 
and $z$ axes are fixed to the separation vector ${\bf r}$ and the Newtonian angular momentum vector ${\bf 
L}_N=\mu{\bf r}\times{\bf v}$, respectively. Here ${\bf v}$ is the relative velocity vector, $\mu=m_1m_2/M^2$ is the 
reduced mass and $M=m_1+m_2$ is the total mass of the system.

The transformation between the coordinate systems is described by Euler-angles \cite{GPV2}. A general vector ${\bf 
u}$ of the comoving system becomes
\begin{eqnarray}
{\bf u}'=R_z(\Phi)R_x(\iota)R_z(\Psi){\bf u}
\end{eqnarray}
in the invariant one. Here $\iota$ is the angle between ${\bf J}$ and ${\bf L}_N$, $\Phi$ describes the precession of 
${\bf L}_N$ over ${\bf J}$ and $\Psi$ represents the direction of the separation vector on the orbital plane, the 
plane perpendicular to ${\bf L}_N$. This way the components of the separation vector in the invariant system are
\begin{eqnarray}
{\bf r}=r\left(\begin{array}{c} \cos{\Phi}\cos{\Psi}-\cos{\iota}\sin{\Phi}\sin{\Psi} \\
\sin{\Phi}\cos{\Psi}+\cos{\iota}\cos{\Phi}\sin{\Psi} \\
\sin{\iota}\sin{\Psi}\end{array} \right)\ .
\end{eqnarray}

With the use of the Euler-angles the relative velocity vector, which is perpendicular to the Newtonian angular 
momentum, is expressed as
\begin{eqnarray}\label{veloc}
{\bf v}=\left(\begin{array}{c} \dot{r} \\
r\left(\cos{\iota}\dot{\Phi}+\dot{\Psi}\right) \\
0 \end{array}\right)
\end{eqnarray}
in the comoving coordinate system. Furthermore we decompose ${\bf v}$ as
\begin{eqnarray}\label{vdec}
{\bf v}=v_{\parallel}{\bf n}+v_{\perp}{\bf m}\ ,
\end{eqnarray}
where ${\bf n}={\bf r}/r$ and ${\bf m}$ is the unit vector parallel to the $y$ axis of the comoving system.

To collect all the terms describing the spin effects of the rotating body we decompose the dynamical quantities into 
zeroth-order and linear terms in spin:
\begin{eqnarray}\label{decomp}
r=r_N+r_S\ ,\qquad
v_{\parallel}=v_{\parallel N}+v_{\parallel S}\ ,\qquad
v_{\perp}=v_{\perp N}+v_{\perp S}\nonumber\\
\Psi=\Psi_N+\Psi_S\ ,\qquad
\Phi=\Phi_N+\Phi_S\ ,\qquad
\iota=\iota_N+\iota_S\ .
\end{eqnarray}

For the calculation of the polarization states we have to determine the components of the orthonormal triad $({\bf 
N},{\bf p},{\bf q})$, where ${\bf N}$ is the direction of the line of sight and ${\bf p}$ is a vector perpendicular 
to ${\bf N}$ and ${\bf L}_N$.

Although the evolution of the angles will be discussed later, we introduce an important result in advance, namely 
$\iota_N=0$ \cite{GPV2}. It is convenient to introduce the angle $\Upsilon=\Psi+\Phi$ which is decomposed into zeroth 
order and linear terms in spin, {\it cf.} Eq. (\ref{decomp}). In this case {\bf N} has the form
\begin{eqnarray}\label{Nvec}
{\bf N}=\left(\begin{array}{c}
\sin{\gamma}\cos{\Upsilon_N}-(\sin{\gamma}\Upsilon_S\sin{\Upsilon_N}- \cos{\gamma}\sin{\Phi_N}\iota_S)\\
-\sin{\gamma}\sin{\Upsilon_N}-(\sin{\gamma}\Upsilon_S\cos{\Upsilon_N}- \cos{\gamma}\cos{\Phi_N}\iota_S)\\
\cos{\gamma}+\sin{\gamma}\sin{\Phi_N}\iota_S
\end{array}\right)
\end{eqnarray}
in the comoving system. Since ${\bf p}$ is a unit vector perpendicular to ${\bf N}$ and ${\bf L}_N$ it becomes
\begin{eqnarray}\label{pvec}
{\bf p}=\left(\begin{array}{c}
\sin{\Upsilon_N}+(\Upsilon_S-\cot{\gamma}\cos{\Phi_N}\iota_S)\cos\Upsilon_N\\
\cos{\Upsilon_N}-(\Upsilon_S-\cot{\gamma}\cos{\Phi_N}\iota_S)\sin\Upsilon_N\\
0
\end{array}\right)
\end{eqnarray}
in the comoving system and ${\bf q}={\bf N}\times{\bf p}$.

\subsection{Spin-precession}

To describe the dynamics of the binary system first we have to determine the evolution of the spin vector ${\bf S}$. 
In the invariant system let the angles $\alpha$ and $\beta$ denote the direction of the spin, 
${\bf S}=S(\sin{\alpha}\cos{\beta},\sin{\alpha}\sin{\beta},\cos{\alpha})$.

The dynamics of ${\bf S}$ is governed by the spin precession equations \cite{BOC}
\begin{eqnarray}\label{Sprec}
\dot{\bf S}=(4+3\zeta)\frac{1}{2r^3}{\bf L}_N\times{\bf S}\ ,
\end{eqnarray}
where $\zeta=m_2/m_1$. Up to 2\,PN order the magnitude $S$ of the spin is constant. Since we are interested in the 
leading order spin effects we may replace ${\bf L}_N\rightarrow{\bf J}$ by inserting higher order terms in $S$. After 
substituting the components of the spin vector Eq. (\ref{Sprec}) leads to the following equations for $\alpha$ and 
$\beta$:
\begin{eqnarray}
\dot{\alpha}=0\ ,\qquad
\dot{\beta}=(4+3\zeta)\frac{J}{2r_N^3}\ .
\end{eqnarray}
The relative PN order of these angles can be determined with the use of $\Upsilon_N$ as zeroth order reference
\begin{eqnarray}
\frac{\dot{\beta}}{\dot{\Upsilon}_N}\sim\frac{L/(r_N^3)}{L/(\mu r_N^2)}=\frac{\mu}{r_N}\sim\epsilon\ ,
\end{eqnarray}
where $\epsilon$ is the post-Newtonian parameter. We assume that the integration of the equations of motion does not 
change the order of the different quantities. Hence $\beta$ can be decomposed as $\beta=\beta_{N}+\beta_{PN}$ with 
${\beta_{N}}/{\Upsilon_N}\sim 1$ and ${\beta_{PN}}/{\Upsilon_N}\sim\epsilon$, and
\begin{eqnarray}
\dot{\beta}_{N}=0\ ,\qquad
\dot{\beta}_{PN}=(4+3\zeta)\frac{J}{2r_N^3}\ .
\end{eqnarray}

\subsection{The equations of motion}

The length and the first component of the relative velocity vector are \cite{GPV2}
\begin{eqnarray}
\!\!\!\!\!\!v^2\!=\!\frac{2E}{\mu}+\frac{2M}{r}-\frac{2\zeta{\bf LS}}{\mu r^3}\ , \quad \!\!
v_{\parallel}^2\!=\!\frac{2E}{\mu}+\frac{2M}{r}-\frac{L^2}{\mu^2r^2}+\frac{2\zeta E{\bf 
LS}}{M\mu^2r^2}-\frac{2(2+\zeta){\bf LS}}{\mu r^3}
\end{eqnarray}
respectively, where $E$ is the energy, $L$ denotes the length of the angular momentum vector ${\bf L}={\bf L}_N+{\bf 
L}_{SO}$ and $\cos{\kappa}={\bf LS}/LS$. These quantities are constants of the motion \cite{Kidder}. Since 
$v_{\perp}^2=v^2-v_{\parallel}^2$, the second component of the relative velocity is
\begin{eqnarray}
v_{\perp}=\frac{L}{\mu r_N}-\frac{L}{\mu r_N}\left(\frac{r_S}{r_N}+\frac{\zeta E{\bf LS}}{L^2M}-\frac{2\mu{\bf 
LS}}{L^2r_N}\right)\ .
\end{eqnarray}
Using Eq. (\ref{veloc}) we get the equations of the motion for the angle $\Upsilon$:
\begin{eqnarray}
\dot{\Upsilon}_N=\frac{L}{\mu r_N^2}\ ,\qquad
\dot{\Upsilon}_S=-\frac{L}{\mu r_N^2}\left(\frac{2r_S}{r_N}+\frac{\zeta E{\bf LS}}{L^2M}-\frac{2\mu{\bf 
LS}}{L^2r_N}\right)\ .
\end{eqnarray}

The remaining equations are determined by the total angular momentum ${\bf J}={\bf L}_N+{\bf L}_{SO}+{\bf S}$,
where
\begin{eqnarray}
{\bf L}_{SO}=\eta\left\{(2+\zeta)\frac{M}{r^3}[{\bf r}\times({\bf r}\times{\bf S})]-\frac{\zeta}{2}[{\bf 
v}\times({\bf v}\times{\bf S})]\right\}
\end{eqnarray}
and $\eta=\mu/M$ \cite{Kidder,GPV2}. After substituting the components of ${\bf r}$, ${\bf v}$ and ${\bf S}$ and 
using the condition that ${\bf J}$ is a constant vector we get the following equations
\begin{eqnarray}
\iota_S\sin{\Phi_N}&=&-(A\cos{\xi}+B\sin{\xi}+C\cos{\beta_N}-\beta_{PN}\sin{\beta_N})\frac{S\sin{\alpha}}{L}\ 
,\nonumber\\
\iota_S\cos{\Phi_N}&=&-(B\cos{\xi}-A\sin{\xi}-C\sin{\beta_N}-\beta_{PN}\cos{\beta_N})\frac{S\sin{\alpha}}{L}\ ,
\end{eqnarray}
where $\xi=2\Upsilon_N-\beta_N$ and
\begin{eqnarray}
A\!=\!\left[\frac{(2\!+\!\zeta)\mu}{2r_N}-\frac{\eta\zeta(v_{\parallel N}^2\!-v_{\perp N}^2)}{4}\right]\!,\, 
B\!=\!\frac{\eta\zeta v_{\parallel N}v_{\perp N}}{2},\,
C\!=\!\left[\frac{\eta\zeta v_N^2}{4}-\frac{(2\!+\!\zeta)\mu}{2r_N}\right].
\end{eqnarray}

Although we cannot give a full description of the angular evolution we have determined all the quantities, namely the 
equations for the angle $\Upsilon$ and the products $\iota_S\cos{\Phi_N}$ and $\iota_S\sin{\Phi_N}$ we need to 
describe the evolution of the polarization states.

\section{The polarization states}

The signal $h(t)$ of a laser-interferometric gravitational wave detector is decomposed into the polarization states 
$h_+(t)$ and $h_{\times}(t)$ \cite{ACST},
\begin{eqnarray}
h(t)=F_+h_+(t)+F_{\times}h_{\times}(t)\ ,
\end{eqnarray}
where $F_+$ and $F_{\times}$ are the so-called beam-pattern functions. The independent polarization states $h_+(t)$ 
and $h_{\times}(t)$ are projected from the transverse-traceless tensor $h^{ij}_{TT}$ representing metric 
perturbations as
\begin{eqnarray}
h_+=\frac{1}{2}(p_ip_j-q_iq_j)h^{ij}_{TT}\ ,\quad
h_{\times}=\frac{1}{2}(p_iq_j+q_ip_j)h^{ij}_{TT} \ .\label{h-k}
\end{eqnarray}
In the post-Newtonian approximation $h^{ij}_{TT}$ can be decomposed as \cite{Kidder}:
\begin{eqnarray}
h^{ij}_{TT}=\frac{2\mu}{D}\left[Q^{ij}+P^{0.5}Q^{ij}+PQ^{ij}+PQ^{ij}_{SO}+P^{1.5}Q^{ij} 
+P^{1.5}Q^{ij}_{SO}\right]_{TT}\ ,\label{hTT}
\end{eqnarray}
where $D$ is the distance between the source and the observer. $Q^{ij}$ denotes the quadrupole (or Newtonian) term, 
$P^{0.5}Q^{ij}$, $PQ^{ij}$ and $P^{1.5}Q^{ij}$ are corrections corresponding to higher PN orders, $PQ^{ij}_{SO}$ and 
$P^{1.5}Q^{ij}_{SO}$ are the spin-orbit terms \cite{Kidder,WW}. Since we are interested in the effects of rotation we 
keep the contributions linear in spin and the quadrupole term. The decomposition of the relative velocity vector, Eq. 
(\ref{vdec}), gives a natural structure to $h^{ij}_{TT}$ and the components can be described in a simple way, see 
Appendix A.

To avoid complicated expressions the components of {\bf N}, {\bf p}, {\bf q}, {\bf v} and {\bf S} are inserted 
formally. We decompose the relevant contributions to the polarization states $h_+$ and $h_{\times}$ as
\begin{eqnarray}
h_{^+_{\times}}=\frac{2\mu}{D}\left[h_{^+_{\times}}{}^N+h_{^+_{\times}}{}^{1SO}+ h_{^+_{\times}}{}^{1.5SO}\right]\ ,
\end{eqnarray}
where
\begin{eqnarray}\label{hpl}
h_+^N&=&\left(\dot{r}^2-\frac{M}{r}\right)(p_x^2-q_x^2)+2v_{\perp}\dot{r}(p_xp_y-q_xq_y)+
v_{\perp}^2(p_y^2-q_y^2)\ ,\nonumber\\
h_+^{1SO}&=&\frac{1+\zeta}{r^2}\left[({\bf qS})p_x+({\bf pS})q_x\right]\ ,\nonumber\\
h_+^{1,5SO}&=&\frac{2}{r^2}\left\{3v_{\perp}(1-\zeta)S_z(p_x^2-q_x^2)+ \frac{(2+\zeta)\dot{r}}{2}[{\bf 
S}\times(p_x{\bf p}-q_x{\bf q})]_x \right.\nonumber\\
&-&\frac{(4-5\zeta)v_{\perp}}{2}[{\bf S}\times(p_x{\bf p}-q_x{\bf q})]_y
-v_{\perp}\zeta[{\bf S}\times(p_y{\bf p}-q_y{\bf q})]_x-\nonumber\\
&-&\left.\zeta{\bf S}\cdot\left[\left(\frac{\dot{r}}{2}N_x+v_{\perp}N_y\right)(p_x{\bf q}+q_x{\bf 
p})+v_{\perp}N_x(p_y{\bf q}+q_y{\bf p})\right]\right\}\ ,
\end{eqnarray}
and similarly
\begin{eqnarray}\label{hcr}
h_{\times}^N&=&2\left(\left[\dot{r}^2-\frac{M}{r}\right]p_xq_x+v_{\perp}\dot{r}(p_xq_y+q_xp_y) 
+v_{\perp}^2p_yq_y\right)\ ,\nonumber\\
h_{\times}^{1SO}&=&\frac{1+\zeta}{r^2}\left[({\bf qS})q_x-({\bf pS})p_x\right]\ ,\nonumber\\
h_{\times}^{1,5SO}&=&\frac{2}{r^2}\left\{6v_{\perp}(1-\zeta)S_zp_xq_x+ \frac{(2+\zeta)\dot{r}}{2}[{\bf 
S}\times(p_x{\bf q}+q_x{\bf p})]_x-\right.\nonumber\\
&-&\frac{(4-5\zeta)v_{\perp}}{2}[{\bf S}\times(p_x{\bf q}+q_x{\bf p})]_y
-v_{\perp}\zeta[{\bf S}\times(p_y{\bf q}+q_y{\bf p})]_x-\nonumber\\
&-&\left.\zeta{\bf S}\cdot\left[\left(\frac{\dot{r}}{2}N_x+v_{\perp}N_y\right)(q_x{\bf q}-p_x{\bf 
p})+v_{\perp}N_x(q_y{\bf q}-p_y{\bf p})\right]\right\}\ .
\end{eqnarray}

Eqs. (\ref{hpl}-\ref{hcr}) form the basics of our main results. With the use of these expressions one can evaluate 
the spin contributions to the polarization states. After parametrizing the orbit \cite{Gop,Param} the equations of 
the motion for the angles $\Upsilon$, $\iota$, $\Phi_0$ and $\beta$ can be integrated. With the substitution of these 
angles into Eqs. (\ref{Nvec}-\ref{pvec}) the explicit parameter dependence of the $({\bf N}$,${\bf p}$,${\bf q})$ 
triad is determined. Using Eqs. (\ref{hpl}-\ref{hcr}) and neglecting quadratic or higher order spin terms one can 
investigate the effects of rotation on the detectable gravitational waveform.

\section{The circular orbit case}

The relevance of the circular orbit case is supported by the fact that gravitational radiation can circularize the 
motion and drive the binary toward the innermost stable circular orbit. Moreover, the dynamics can be integrated 
explicitly in time in this limit. Although the main steps of the method given above do not change some equations and 
expressions become simpler.

The relative velocity vector can be generally decomposed as 
\begin{eqnarray}
{\bf v}=\dot{r}{\bf n}+r\omega{\bf m}
\end{eqnarray}
and circular orbits are defined by the $\dot{r}=0$ and $\dot{\omega}=0$ conditions. To obtain the expressions for the 
polarization states we use the relevant terms in the decomposition of the relative velocity given in Ref. 
\cite{Kidder}:
\begin{eqnarray}
v^2=v_{\perp}^2=r^2\omega^2=\frac{M}{r}\left[1-\frac{1}{M^2}(2+3\zeta)({\bf n}\times{\bf m}){\bf 
S}\left(\frac{M}{r}\right)^{3/2}\right]\ .
\end{eqnarray}
Then the contributions for $h_+$ and $h_{\times}$ which change compared to the general formula become
\begin{eqnarray}
\!\!\!h_+^N\!&=&\!\left(\frac{M}{r}\right)[-(p_x^2-q_x^2)+(p_y^2-q_y^2)]\ ,\nonumber\\
\!\!\!h_+^{1,5SO}\!&=&\!\frac{2v}{r^2}\left[S_z(p_x^2-q_x^2)+2S_x(p_xp_z-q_xq_z)\right]+\nonumber\\ 
&+&\!\frac{v\zeta}{r^2}\left[S_z(p_x^2-q_x^2)+2S_z(p_xp_y-q_xq_y)+5S_x(p_xp_z-q_xq_z)+\right.\nonumber\\
&+&\!\left.2S_x(p_yp_z-q_yq_z)\!-\!2N_y(({\bf qS})p_x\!+\!({\bf pS})q_x)\!-\!2N_x(({\bf qS})p_y\!+\!({\bf 
pS})q_y)\right]
\end{eqnarray}
and
\begin{eqnarray}
\!\!\!h_{\times}^N\!&=&\!2\left(\frac{M}{r}\right)[-p_xq_x+p_yq_y]\ ,\nonumber\\
\!\!\!h_{\times}^{1,5SO}\!&=&\!\frac{4v}{r^2}\left[S_zp_xq_x+S_x(p_xq_z+q_xp_z)\right]+\nonumber\\ 
&+&\!\frac{v\zeta}{r^2}\left[2S_zp_xq_x-2S_z(p_xq_y+q_xp_y)+5S_x(p_xq_z+q_xp_z)+\right.\nonumber\\
&+&\!\left.2S_x(p_yq_z+q_yp_z)\!-\!2N_y(({\bf qS})q_x\!-\!({\bf pS})p_x)\!-\!2N_x(({\bf qS})q_y\!-\!({\bf 
pS})p_y)\right]\!\!\ .
\end{eqnarray}

\section{Extreme mass ratio limit}

An important special case discussed in the literature is the extreme mass ratio limit, when the mass ratio of the 
bodies is negligible. We consider the case when the mass of the rotating body is much greater than the other. This 
way $\eta\approx\zeta\ll 1$, $M\approx m_1$, $\mu\approx m_2$ and $\delta m/m\approx 1$ and the description of the 
motion and the formal expressions for the polarization states change.

In the extreme mass ratio limit the components of the relative velocity become
\begin{eqnarray}
v^2&=&\frac{2E}{m_2}+\frac{2m_1}{r}\ ,\qquad
\dot{r}^2=\frac{2E}{m_2}+\frac{2m_1}{r}-\frac{L^2}{m_2^2r^2}-\frac{4{\bf LS}}{m_2r^3}\ ,\nonumber\\
v_{\perp}&=&\frac{L}{m_2r_N}-\frac{Lr_S}{m_2r_N^2}+\frac{2{\bf LS}}{Lr_N^2}\ ,
\end{eqnarray}
and the equations of the motion for the angle variables are:
\begin{eqnarray}
\dot{\beta}_{PN}&=&\frac{2J}{r_N^3}\ ,\qquad
\dot{\Upsilon}_N=\frac{L}{m_2r_N^2}\ ,\qquad
\dot{\Upsilon}_S=-\frac{2Lr_S}{m_2r_N^3}+\frac{2{\bf LS}}{Lr_N^3}\ ,\nonumber\\
\iota_S\sin{\Phi_N}&=&-\left(\frac{m_2}{r_N}\cos{\xi}- 
\frac{m_2}{r_N}\cos{\beta_N}-\sin{\beta_N}\beta_{PN}\right)\frac{\sin{\alpha}S}{L}\ ,\nonumber\\
\iota_S\cos{\Phi_N}&=&\left(\frac{m_2}{r_N}\sin{\xi} 
-\frac{m_2}{r_N}\sin{\beta_N}+\cos{\beta_N}\beta_{PN}\right)\frac{\sin{\alpha}S}{L}\ .
\end{eqnarray}
The contributions for the polarization states can be written as
\begin{eqnarray}
\!\!\!h_+^N\!&=&\!\left(\dot{r}^2-\frac{m_1}{r}\right)(p_x^2-q_x^2)+2v_{\perp}\dot{r}(p_xp_y-q_xq_y)+ 
v_{\perp}^2(p_y^2-q_y^2)\ ,\nonumber\\
\!\!\!h_+^{1SO}\!&=&\!\frac{1}{r^2}\left[({\bf qS})p_x+({\bf pS})q_x\right]\ ,\nonumber\\
\!\!\!h_+^{1,5SO}\!&=&\!\frac{2}{r^2}\left\{3v_{\perp}S_z(p_x^2-q_x^2)\!+\!\dot{r}[{\bf S}\times\!(p_x{\bf 
p}\!-\!q_x{\bf q})]_x\!-\!2v_{\perp}[{\bf S}\times\!(p_x{\bf p}\!-\!q_x{\bf q})]_y\right\}\!\!\ ,
\end{eqnarray}
and similarly
\begin{eqnarray}
\!\!\!h_{\times}^N\!&=&\!2\left(\left[\dot{r}^2-\frac{m_1}{r}\right]p_xq_x+v_{\perp}\dot{r}(p_xq_y+q_xp_y) 
+v_{\perp}^2p_yq_y\right)\ ,\nonumber\\
\!\!\!h_{\times}^{1SO}\!&=&\!\frac{1}{r^2}\left[({\bf qS})q_x-({\bf pS})p_x\right]\ ,\nonumber\\
\!\!\!h_{\times}^{1,5SO}\!&=&\!\frac{2}{r^2}\left\{6v_{\perp}S_zp_xq_x\!+\!\dot{r}[{\bf S}\times\!(p_x{\bf 
q}\!+\!q_x{\bf p})]_x\!-\!2v_{\perp}[{\bf S}\times\!(p_x{\bf q}\!+\!q_x{\bf p})]_y\right\}\!\!\ .
\end{eqnarray}

In a previous work \cite{LT1} the polarization states was described in the Lense-Thirring approximation. The $z$ axis 
of the invariant system was fixed to ${\bf S}$ since the precession of the spin is negligible in that case. There is 
a constant rotational transformation between the above description and the Lense-Thirring case. The main difference 
is that in the Lense-Thirring approximation $\dot{\bf J}$ fails to be zero. If one specifies that $\dot{\bf J}=0$ in 
this comparable mass case the common limit of the two descriptions can be found.

\section{Conclusions and remarks}

In this article we have presented a method to evaluate the detectable gravitational wave signals generated by a 
spinning compact binary system moving on eccentric orbit in the case of one spinning object up to 1.5\,PN order. We 
have introduced the invariant and a comoving coordinate systems to describe the evolution of the dynamical 
quantities. With the use of the constants of motion we have discussed the equations describing the evolution of the 
dynamical quantities for the determination of the polarization states. We have calculated the components of the 
relative velocity vector, the spin and the $({\bf N}$,${\bf p}$,${\bf q})$ triad in terms of these quantities, namely 
the length of the separation vector and the Euler-angles. To determine the effect of the eccentricity of the orbit on 
the detectable signals we have investigated the circular orbit limit. An other significant property of this case is 
that the explicit time dependence of $h_+$ and $h_{\times}$ can be calculated. 

The results presented here are independent of the parameterization of the orbit. We plan to use the generalized true 
anomaly parameterization of the motion \cite{Param} to investigate the structure of the wave signals. Moreover, this 
method can be the starting point to study the properties of two spinning objects and unbound orbits \cite{KGM}.

\section*{Acknowledgments}
This work was supported by OTKA no. TS044665, F049429 and T046939 grants.

\appendix

\section{The transverse-traceless tensor}

The decomposition of the relative velocity vector, Eq. (\ref{vdec}), results the following form of the quadrupole and 
spin-orbit terms of $h^{ij}_{TT}$:
\begin{eqnarray}
Q^{ij}&=&2\left(\left[\dot{r}^2-\frac{M}{r}\right]n^in^j+2v_{\perp}\dot{r}m^{(i}n^{j)}+ v_{\perp}^2m^im^j\right),\\
PQ^{ij}_{SO}&=&-\frac{2M}{r^2m_1}[{\bf S}\times{\bf N}]^{(i}{\bf n}^{j)},\\
P^{1,5}Q^{ij}_{SO}&=&\frac{2}{m_1r^2}\left(6v_{\perp}M[{\bf n}\times{\bf m}]{\bf S}n^in^j+ 
(2m_1-m_2)\dot{r}n^{(i}[{\bf n}\times{\bf S}]^{j)}\right.\nonumber\\
&-&(5m_2+4m_1)v_{\perp}n^{(i}[{\bf m}\times{\bf S}]^{j)}-2m_2v_{\perp}m^{(i}[{\bf m}\times{\bf S}]^{j)}\nonumber\\
&-&\left.(\dot{r}N_x+2v_{\perp}N_y)m_2[{\bf S}\times{\bf N}]^{(i}n^{j)}-2N_xv_{\perp}m_2[{\bf S}\times{\bf 
N}]^{(i}m^{j)}\right).
\end{eqnarray}

\end{document}